\documentclass[aps,twocolumn,pra,superscriptaddress,amsmath,showpacs,tightenlines]{revtex4-1}
\usepackage{amssymb}
\usepackage{amsmath}
\usepackage{graphicx}
\usepackage{subfigure}
\usepackage{natbib}
\usepackage{epsfig}
\usepackage{amsfonts}
\usepackage{mathrsfs}
\usepackage{CJK}
\usepackage{xcolor}
\usepackage[toc,page,title,titletoc,header]{appendix}

\begin{document}
\title{{Phase-controlled  bound states in giant atom waveguide QED setup} }
\author{Xiaojun Zhang}
\affiliation{Center for Quantum Sciences and School of Physics, Northeast Normal University,
Changchun 130024, China}
\author{Mingjie Zhu}
\affiliation{Center for Quantum Sciences and School of Physics, Northeast Normal University,
Changchun 130024, China}
\author{Zhihai Wang}
\email{wangzh761@nenu.edu.cn}
\affiliation{Center for Quantum Sciences and School of Physics, Northeast Normal University,
Changchun 130024, China}

\begin{abstract}
{We find the phase-controlled bound state out of the continuum~(BOC) and bound state in the continuum~(BIC) in an artificial giant atom dressed one-dimensional photonic waveguide where the giant atom couples to the waveguide via two distant sites. We obtain the existence condition of the BOC as well as the frequency and the photonic distribution in the BIC. More interestingly, we predict the quantum beats in the atomic and photonic dynamical evolution, which is induced by the oscillation between the BIC and the BOCs. These findings provide an approach to manipulate the waveguide system via the bound states.}
\end{abstract}
\maketitle

\section{introduction}
The waveguide quantum electrodynamics (QED) investigates the light-matter interaction in the confined waveguide structures. It plays an important role in constructing quantum network and performing kinds of quantum information processing. In the traditional waveguide community, the natural atom or artificial atom was usually considered as point like dipole. However, in 2014, the coherent coupling between the superconducting transmon qubit and the surface acoustic waves~\cite{S1} break down the dipole approximation. In this setup, the size of the transmon is in the same order of the acoustic wavelength, and one has to consider the effects of non-local coupling via multiple points, leading to giant atom quantum optics. Subsequently, the giant atom setup was realized by coupling the transmon or magnon spin ensemble to the curved transmission line and proposed in the synthesis dimension domain~\cite{sq1,sq2,sq3,mq,pro1,pro2,pro3}. The interference and retardation effect due to the non-local coupling in the giant atom system has evoked lots of attentions theoretically to demonstrate some exotic phenomena, such as non-Markovian oscillation~\cite{NM1,NM2}, decoherence-free interaction~\cite{AF2018,AC2020}, retardation effect~\cite{chengre,LG2017}, Lamb shift~\cite{Lambpra14,Lambpra23} and chiral photonic population~\cite{XW2021,XW2022,AS2022,FR2024}. The combination of the topological photons and the giant atom  also induces some interesting effects, for example the chiral zero mode~\cite{chengre,tp2}, the vacuum like dressed states~\cite{vds,sc}, the wide band photonic reflection~\cite{tpg}.

The waveguide provides a quantum channel for the propagation of the photon, which is the ideal carrier of the quantum information. Meanwhile, the waveguide supplies a structured environment for the manipulation of quantum emitters. One of the promising topics in waveguide QED is the bound states,  which are dressed by both of the atom and the photon. The conventional knowledge tells us that the bound state was always far away from the energy band in frequency. Such bound state out of the continuum (BOC)~\cite{ZBOC1,Peter16pra} had potential applications in preserving quantum coherence~\cite{CJ2016} and entanglement~\cite{QJ2010}, realizing quantum precise measurement~\cite{KB2019,liu} and controlling the atomic evolution and photonic propagating~\cite{BS1,BS2,Qiao2021,ZLpra08,BOC3,BOC2}.  Recently,  the bound state in the continuum (BIC)~\cite{BIC1,BIC2,BIC3}, in which the frequency is inside the propagation band, is attracting more and more attentions due to their applications in enhancing light-matter interaction~\cite{BIC6,bicre1,bicre2}, generating low threshold laser~\cite{laser1,laser2,laser3}, enhancing the nonlinear responds~\cite{nonlinear1,nonlinear2}, realizing the quantum sensing~\cite{sense1,sense2} and high efficient wave guiding~\cite{waveguiding1,waveguiding2,waveguiding3, waveguiding4}, etc.

Nowadays, the coupled resonator waveguide (CRW) has been fabricated on demand in the superconducting circuits and is characterized by its good expansibility and integration~\cite{CRW1,CRW2,CRW3}. The cosine type dispersion relation of the CRW brings us a new chance to manipulate the light-matter interaction. Especially, the BOC and BIC in the atom-CRW coupled system has attracted lots of interests recently~\cite{BIC7,ZZ,SL2021,Qiao2021,Peter16pra,BOC2}. However, most of these studies the character of the bound state, how to control their frequency and the wave function are still lack of discussions.

To tackle the above issues, we here promote a scheme to realize the BOC and BIC engineering by coupling a giant atom to the coupled resonator waveguide via two sites non-locally. Comparing with our previous work in Refs.~\cite{BOC2,ZZ}, we here focus on how to control the BOC and BIC via the coupling phase between the giant atom and the CRW. We analytically obtain the conditions for the existence of the BOCs, the eigen frequency and the photonic distribution for the BIC. Moreover, we find that the BIC-BOC oscillation leads to the exotic quantum beat phenomenon, which does not exist in the tradition small atom setup and giant atom setup whose frequency is resonant with the bare resonator in the waveguide~\cite{ZZ}.

\section{model}
\begin{figure}[t]
  \centering
  \includegraphics[width=8cm]{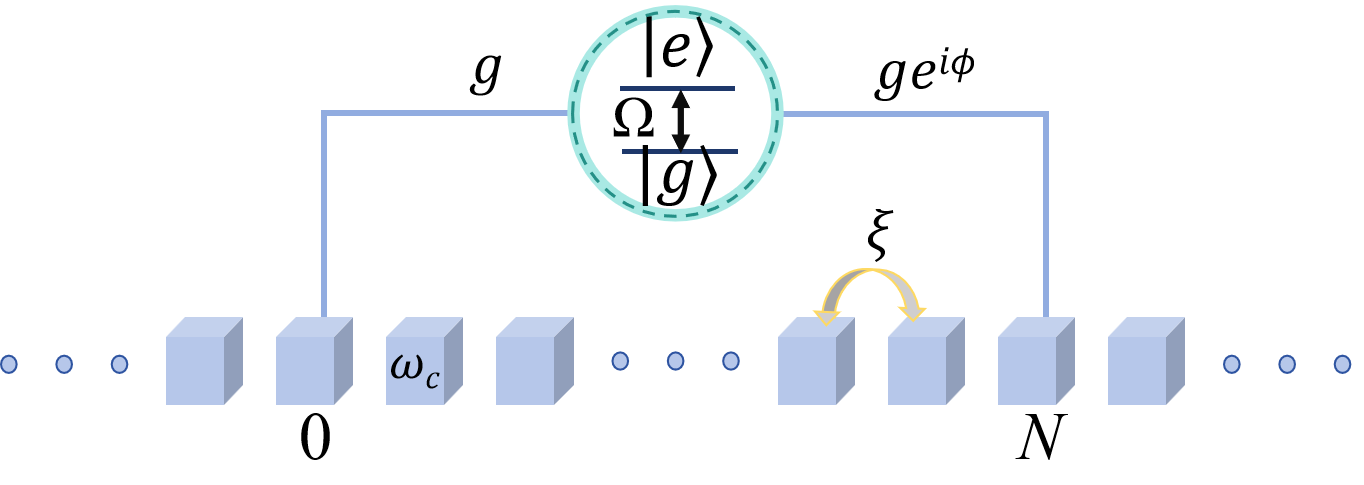}\\
  \caption{Sketch of the waveguide QED setup, where a giant atom
is coupled to a coupled resonator waveguide via the $0$th and the $N$th sites.}\label{schem}
\end{figure}

As schematically shown in Fig.~\ref{schem}, a single giant atom is coupled to the CRW via the $0$th and $N$th sites. The Hamiltonian of the structure in real space reads $H=H_A+H_c+H_I$ where  $(\hbar=1)$
\begin{eqnarray}
H_A&=&\Omega |e\rangle\langle e|,\\
H_c&=&\omega_c \sum_{j}a_{j}^{\dagger}a_{j}-\xi \sum_{j}(a_{j+1}^{\dagger}a_{j}+a_{j}^{\dagger}a_{j+1}),\\
H_I&=&g[(a_{0}^{\dagger}+a_{N}^{\dagger}e^{i\phi})\sigma_- +{\rm H.C.} ].
\end{eqnarray}
Here, $H_A$ and $H_c$ are the Hamiltonian of the giant atom and the CRW respectively and $H_I$ represents their interaction. Here we have performed the rotating wave and dipole approximations at each coupling point between the giant atom and the CRW. $\Omega$ is the transition frequency of the giant atom between the ground state $|g\rangle$ and the excited state $|e\rangle$. $\omega_c$ is the intrinsic frequency of each resonator in the CRW.  $a_j$ is the annihilation operator of the $j$th resonator in the CRW, $\sigma_+=|e\rangle\langle g|$ is the raising operator of the giant atom. $\xi$ is the hopping strength between the nearest resonators in the CRW. $g$ is the real coupling strength between the giant atom and the connected resonator. We have assumed that the coupling phase in the left leg is zero while that for the right leg is $\phi$. {For the convenience of the calculation, we will shift the energy level of the Hamiltonian and set $\omega_c$ to be $0$ in what follows.}

We consider that the CRW is composed by infinite resonators so that the waveguide satisfies the translation invariance. By the Fourier transformation $a^{\dagger}_{k}=\sum_{j}a^{\dagger}_{j}\exp(-ikj)/\sqrt{N_c}$ with $N_c\rightarrow\infty$ being the length of the waveguide, {the Hamiltonian of the CRW $H_c$ is expressed as $H_c=\sum_{k}\omega_k a_{k}^{\dagger}a_{k}$ where the dispersion relation satisfies $\omega_k = -2\xi \cos k$. It describes that the single photon energy band possesses a width of $4\xi$.} In term of the operator $a_k$, the interaction Hamiltonian becomes
\begin{eqnarray}
H_I&=&\frac{g}{\sqrt{N_c}}\sum_{k}[ a_{k}^{\dagger}\sigma_{-}(1+e^{i(kN+\phi)})+ {\rm H.c.}],\label{Hk}
\end{eqnarray}
It is obvious that the coupling amplitude between the giant atom and its resonate mode in the CRW is
\begin{equation}
G(\phi)=\frac{g}{\sqrt{N_c}}(1+e^{i(KN+\phi)})
\end{equation}
which is $\phi$ dependent and $K=-\arccos (\Omega/2\xi)$.

\begin{figure}
  \includegraphics[width=4.1cm]{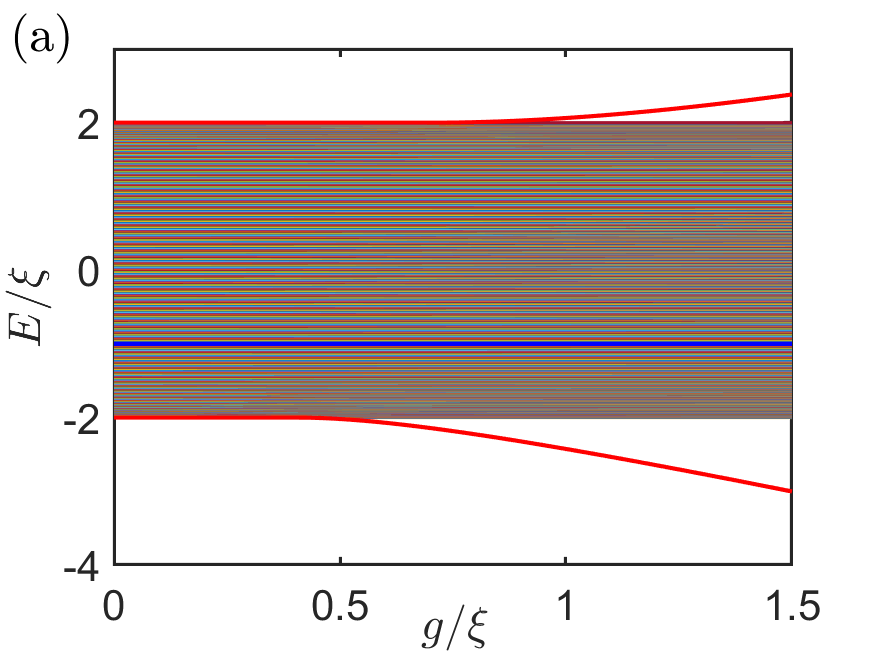}
  \includegraphics[width=4.1cm]{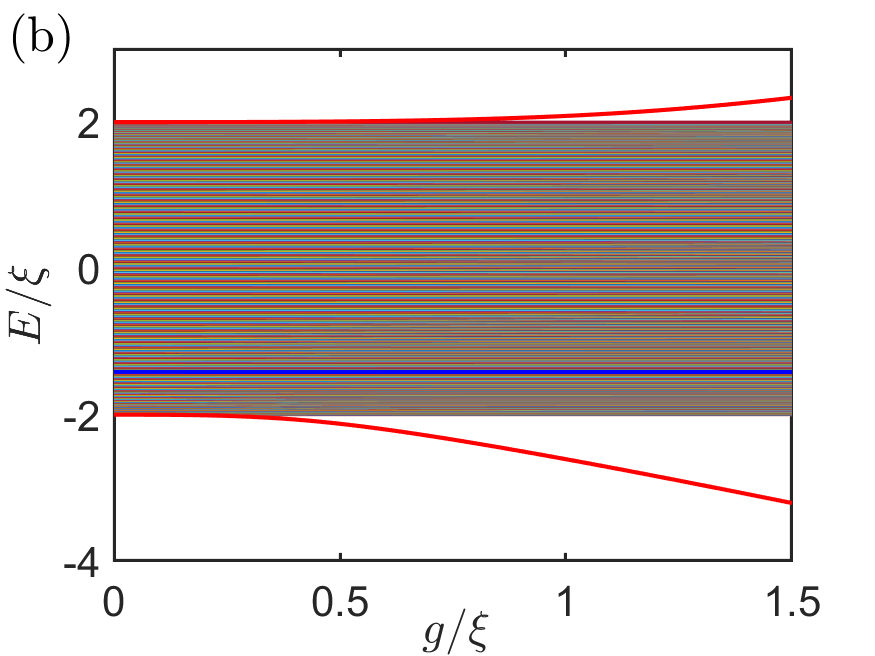}
  \caption{The energy diagram of the giant atom-CRW coupled system where the thick blue line in the continuum is the BIC, while the red curves are the BOC. The parameters are set as  $\Omega=-\xi,N=6,\phi=\pi$ in (a) and $\Omega=-\sqrt{2}\xi,N=12,\phi=0$ in (b).} \label{cc}
\end{figure}

{In Figs.~\ref{cc}(a) and (b), we numerically plot the energy spectrum of system in the single excitation subspace for different atomic resonant frequencies and coupling phases $\phi$. In both of the parameter conditions, we find a pair of states which locate above and below the continual band respectively, as demoted by the red solid curves. Obviously, these two states are BOCs, since they are separated from the continuum. They originate from the broken of the translational symmetry of the waveguide induced by the giant atom. The BOCs also exist in the CRW system which couples to the traditional small atom(s)~\cite{Peter16pra,ZLpra08,Qiao2021}. Furthermore, we also find a single BIC, which is labelled by the blue line in the figure. The BIC is resonant with the frequency of the giant atom, and originates from the destructive coupling between the two atom-waveguide coupling site. For the detailed discussions about why they are ``bound states'', see what follows.}

{To investigate these states deeply, we resort to the single excitation wave function ansatz of
\begin{equation}
\left|\phi\right\rangle= \alpha\sigma_+|G\rangle+\sum_{k}\beta_{k}a_{k}^{\dagger}\left|G\right\rangle,\label{wavek}
\end{equation}
 where $|G\rangle$ represents that both the giant atom and the waveguide are in their ground states. According to the Schr\"odinger equation $H|\phi\rangle=E|\phi\rangle$, we can get
{\begin{eqnarray}
E-\Omega&=&Y(E), \label{BIC}\\
\frac{\beta_k}{\alpha}&=&\frac{1}{\sqrt{N_c}}\frac{[g+ge^{i(kN+\phi)}]}{E+2\xi\cos k},
\end{eqnarray}}}
where the auxiliary function $Y(E)$ is defined as
\begin{equation}
Y(E)\equiv\frac{g^2}{\pi}\int_{-\pi}^{\pi}\frac{1+\cos(kN+\phi)}{E+2\xi\cos k}dk.
\end{equation}

\section{Bound states out of the continuum}

In this section, we will investigate the BOCs of the system by solving the energy equation in Eq.~(\ref{BIC}) in the regime of $|E|>2\xi$. For simplicity, we denote the topmost and bottommost states (eigen energies) as $|E_U\rangle\,(E_U)$ and $|E_L\rangle\,(E_L)$, respectively.  After the detailed calculations as shown in Appendix~\ref{ABOC}, we find that the corresponding energy satisfies the transcendental equation $E_{U(L)}-\Omega=Y(E_{U(L)})$, where
{\begin{eqnarray}
Y(E_U)&=&\frac{g^2}{\xi}\frac{1+[-\frac{E_U}{2\xi}
+\sqrt{(\frac{E_U}{2\xi})^2-1}]^N\cos\phi}{\sqrt{(\frac{E_U}
{2\xi})^2-1}},\label{EUm}\\
Y(E_L)&=&-\frac{g^2}{\xi}\frac{1+[-\frac{E_L}{2\xi}-
\sqrt{(\frac{E_L}{2\xi})^2-1}]^N\cos\phi}{\sqrt{(\frac{E_L}{2\xi})
^2-1}}.\label{ELm}
\end{eqnarray}}
$Y_{E_{U}}$ and $Y_{E_{L}}$ are the monotone {decreasing} function in the regime of $E_{U}>2\xi$ and $E_L<-2\xi$, respectively. Therefore, the conditions for the existence of the upper and lower BOCs can be written as~\cite{BOC3}
\begin{eqnarray}
\lim_{E_U\rightarrow 2\xi}Y_U(E_U)&>&2\xi-\Omega\label{EU1},\\
\lim_{E_L\rightarrow -2\xi}Y_L(E_L)&<&-2\xi-\Omega\label{EL1}.
\end{eqnarray}
In Table~I, we list the above conditions, focusing on the atom-waveguide coupling strength $g$ with different $\phi$. Here, we use the notation ``$\surd$'' to denote that the corresponding BOC always exists regardless of the atom-waveguide coupling strength.

\begin{table}
 \centering
 {\begin{tabular}{|c|c|c|c|}
  \hline
  $E_U$& $\phi=(2n+1)\pi$ & $\phi=2n\pi$ & $\phi=$other  \\
  \cline{1-4}
  $N=2n+1$ & $\surd$ & $g^2>\frac{2\xi^2-\Omega\xi}{N}$ & $\surd$\\
   \cline{1-4}
  $N=2n$ &  $g^2>\frac{2\xi^2-\Omega\xi}{N}$ & $\surd$& $\surd$ \\
  \hline
  \hline
   $E_L$& $\phi=(2n+1)\pi$ & $\phi=2n\pi$ & $\phi=$other  \\
  \cline{1-4}
  $N=2n+1$ & $g^2>\frac{2\xi^2+\Omega\xi}{N}$& $\surd$ & $\surd$\\
   \cline{1-4}
  $N=2n$ &  $g^2>\frac{2\xi^2+\Omega\xi}{N}$ & $\surd$& $\surd$ \\
  \hline
\end{tabular}
\caption{The condition for the existence of the bound state. Here, $n$ is taken to be integer. $\surd$ represents that the corresponding BOC always exists regardless of the atom-waveguide coupling strength $g$.}\label{g}}
\end{table}

\begin{figure}
  \includegraphics[width=4cm]{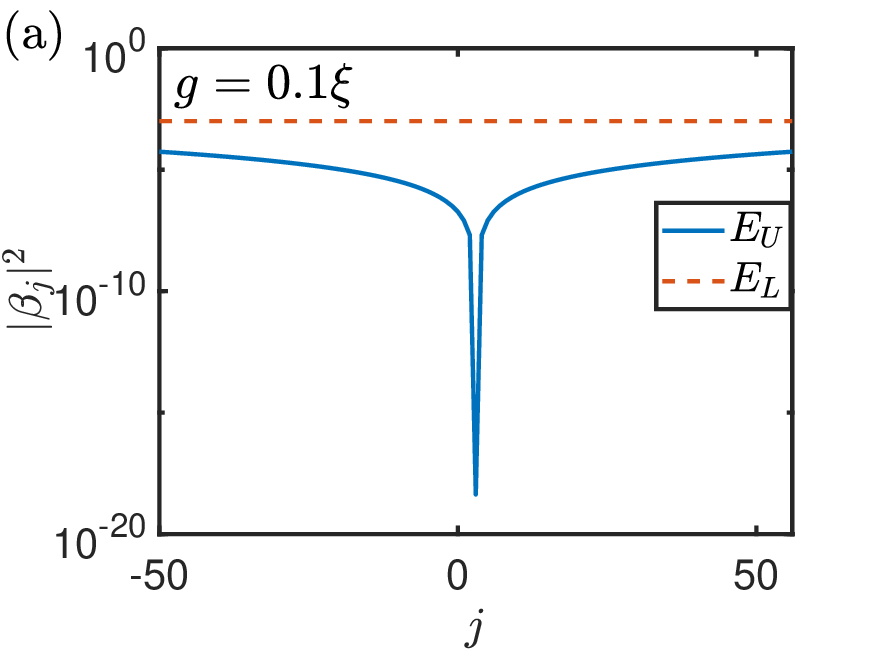}\includegraphics[width=4cm]{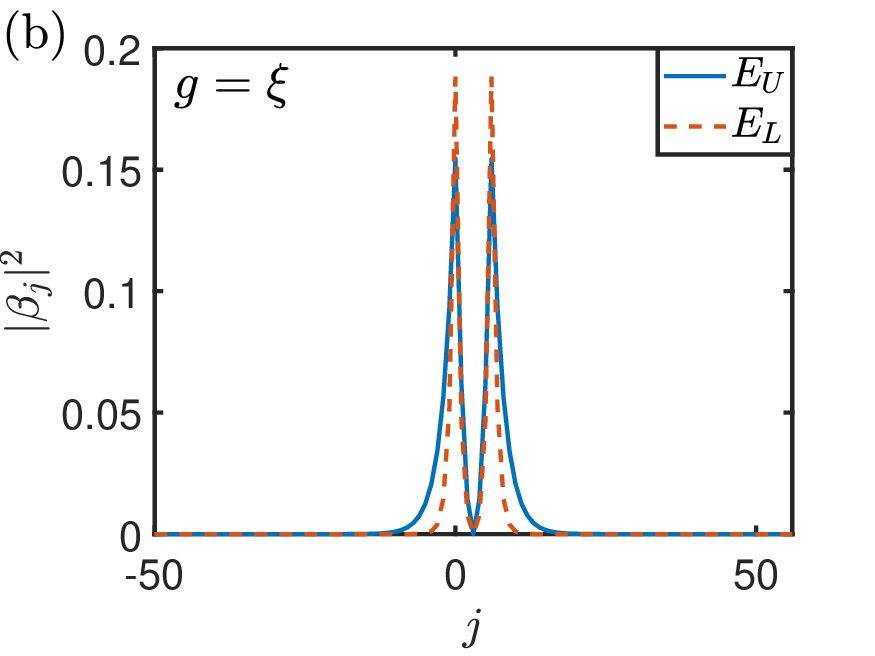}\\
  \includegraphics[width=4cm]{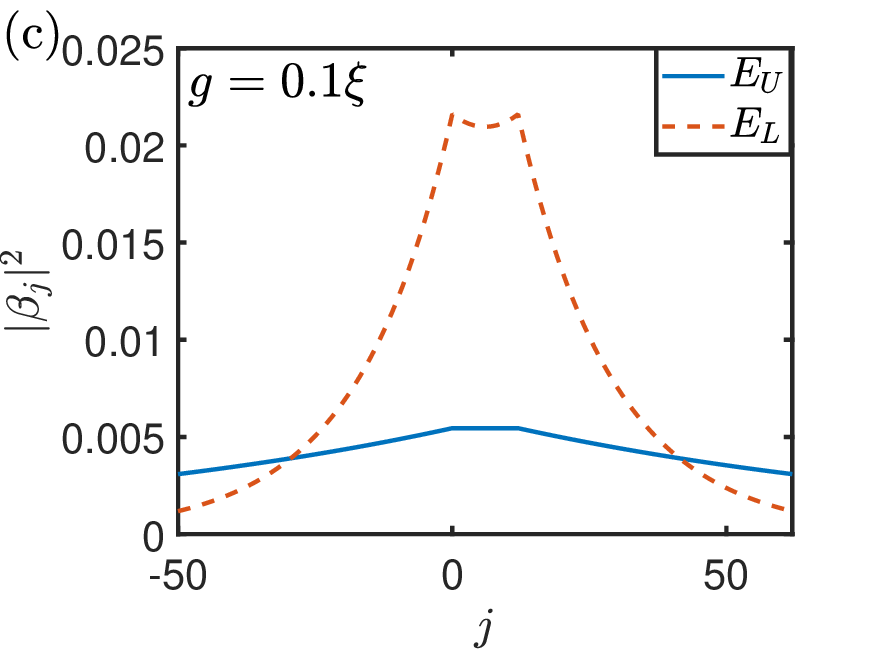}\includegraphics[width=4cm]{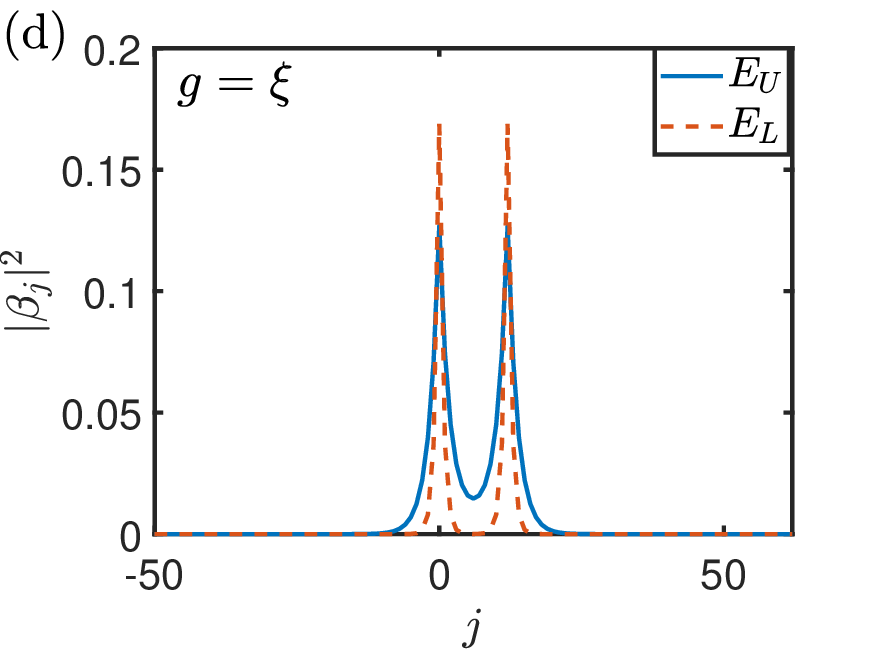}\\
  \includegraphics[width=4cm]{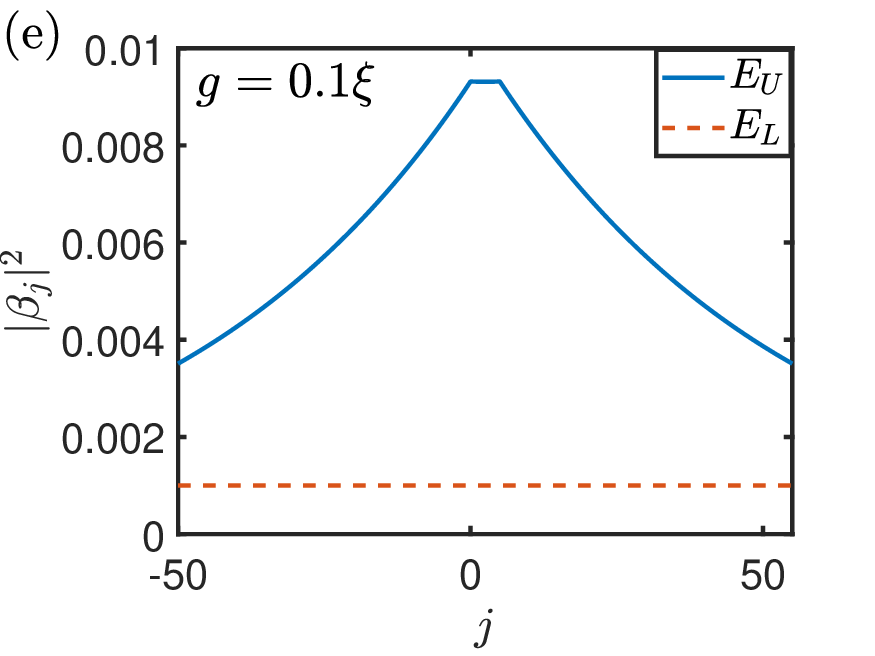}\includegraphics[width=4cm]{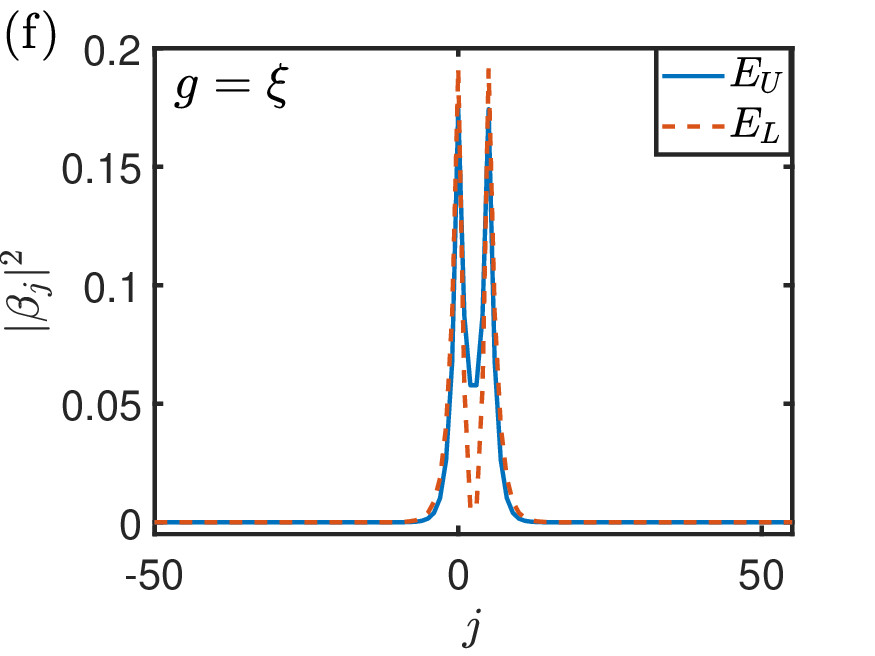}\\
  \includegraphics[width=4cm]{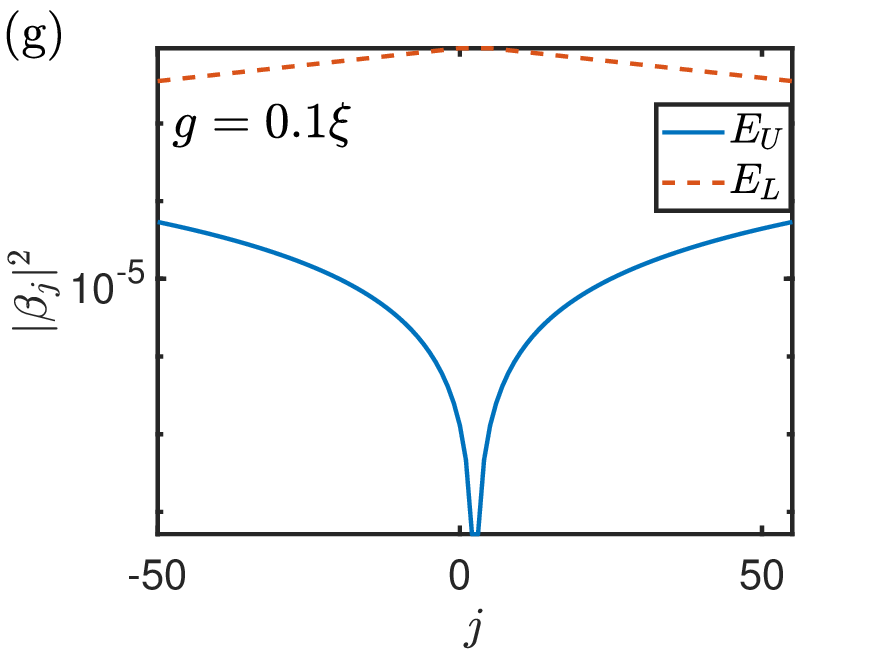}\includegraphics[width=4cm]{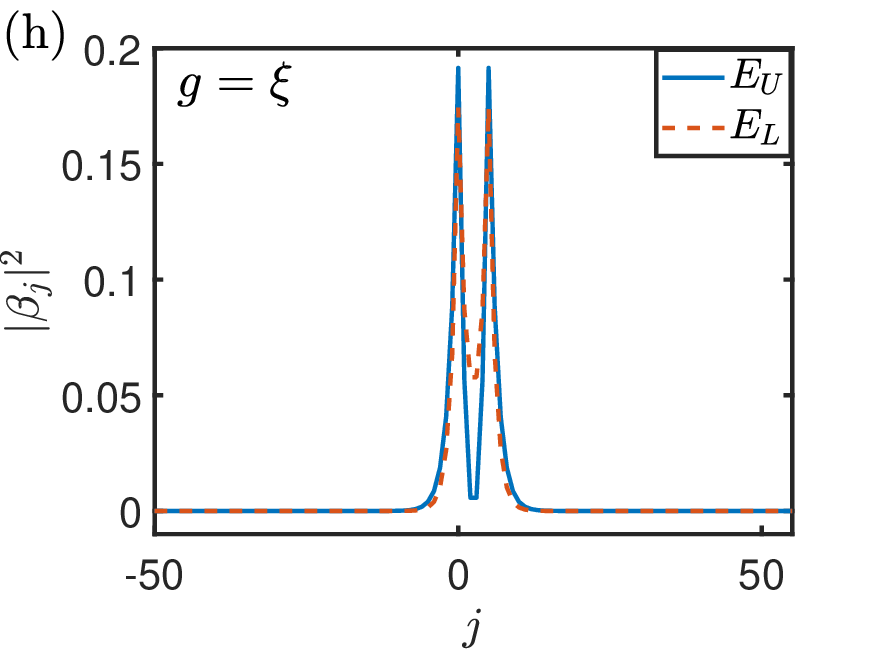}
\caption{The photonic  population in the waveguide for the BOCs $|E_U\rangle$ and $|E_L\rangle$. The parameters are set as (a,b) $\Omega=-\xi,N=6,\phi=\pi$. (c,d) $\Omega=-\sqrt{2}\xi,N=12,\phi=0$. (e,f) $\Omega=0,N=5,\phi=\pi$. (g,h) $\Omega=0,N=5,\phi=0$.}\label{pt}
\end{figure}

{Based on the conditions listed in Table \ref{g}, we further check the photonic distribution for the topmost and bottommost state in Fig.~\ref{cc} (a) and (b) (both for even $N$), respectively. For the small coupling strength $g=0.1\xi$, both of the $E_U$ and $E_L$ exhibit an extended character in Fig.~\ref{pt}(a) for $\phi=\pi$, implying that both of them are not the bound states. On the contrary, the result for $\phi=0$ in Fig.~\ref{pt}(c) shows that the photonic amplitudes decrease slowly outside the regime covered by the giant atom, which is a typical feature of the bound state. To achieve a tighter bound state, we plot the results for larger $g$ in Figs.~\ref{pt}(b) and (d), in which the photon is nearly bounded at the two atom-waveguide coupling sites, for both of $\phi=0$ and $\phi=\pi$. Furthermore, we demonstrate the results for odd $N$ in Fig.~\ref{pt} (e-h). {When $\phi=\pi$, we observe that the topmost state is the bound state, but the bottommost one is not, as shown in Figs.~\ref{pt}(e). When $\phi=0$, the result is completely opposite to that of $\phi=\pi$, as shown in Fig.~\ref{pt}(g) . } As for the strong atom-waveguide coupling, the photonic distribution in the two states in Figs.~\ref{pt}(f) and (h) shows that both of them are well-defined bound states, which are similar to those for even $N$. These results are consistent with the conditions summarized in Table \ref{g}. }

\section{Bound state in the continuum}

\begin{figure}
 \includegraphics[width=4.1cm]{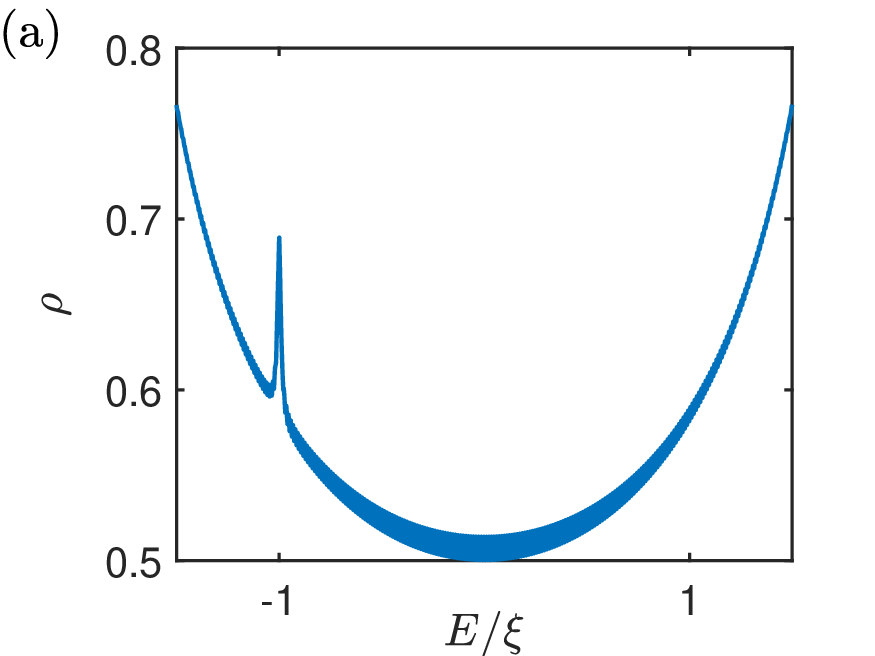}\includegraphics[width=4.1cm]{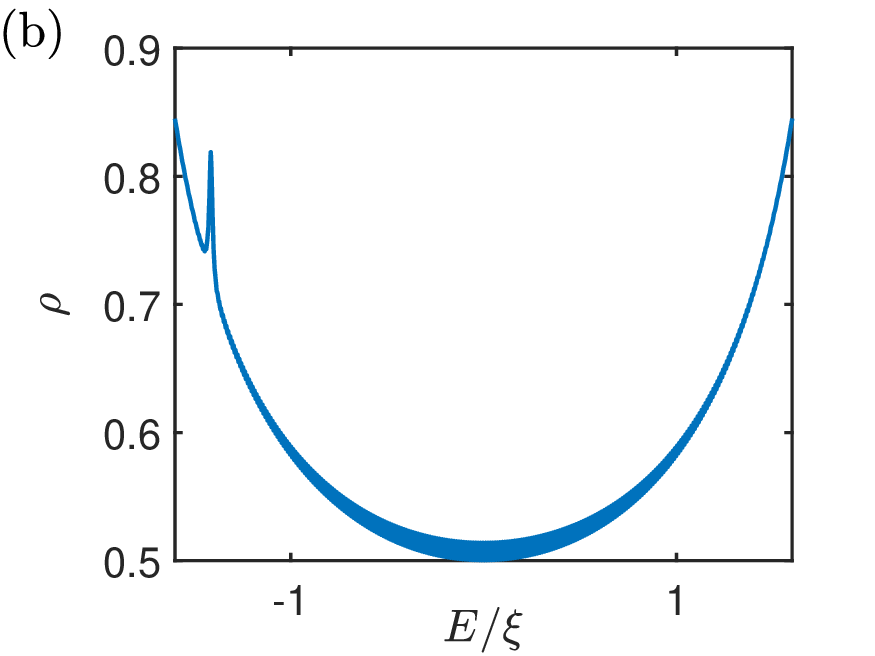}\\
 \includegraphics[width=4.1cm]{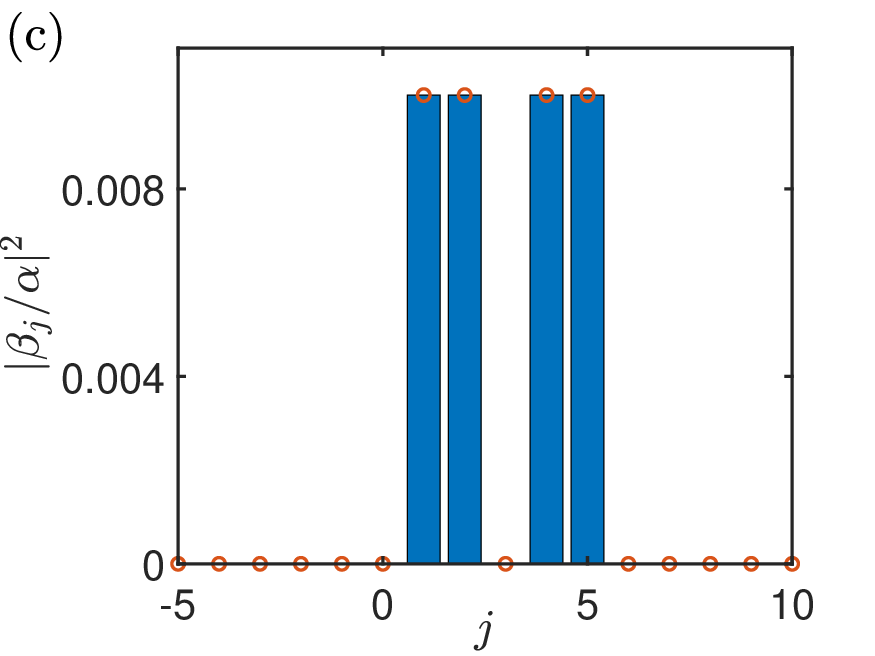}\includegraphics[width=4.1cm]{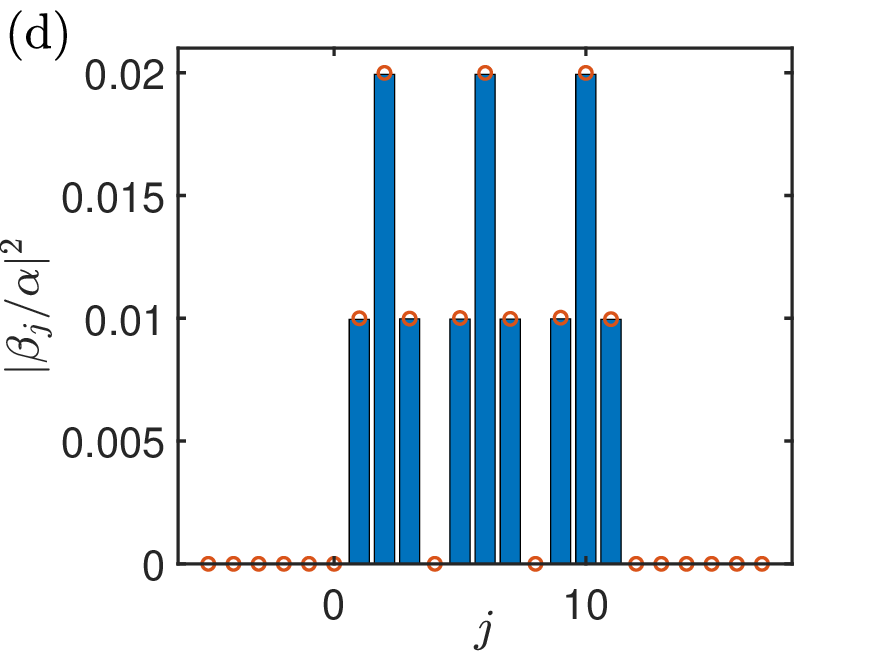}
  \caption{ The density of state (a,b) and the photonic distribution (c,d) of the giant atom-waveguide coupled system. The parameters are set as  $g=0.1\xi,\omega_c=0$ and $\Omega=-\xi,N=6,\phi=\pi$ in (a,c), and $\Omega=-\sqrt{2}\xi,N=12,\phi=0$ in (b,d).}\label{bshape}
\end{figure}

In the last section, we have discussed the BOCs in the giant atom-waveguide coupled system. Now, we move to the continual band. In Fig.~\ref{bshape} (a) and (b), we plot the density of state (Dos) $\rho$ in the regime of $-2\xi<E<2\xi$ when the giant atom is decoupled from its resonant mode in the waveguide, that is, $|G(\phi)|^2=0$. We observe a single peak in both of the figures, which locate exactly at the atomic frequency $\Omega$. As derived in Appendix~\ref{appB}, we can prove that $E=\Omega$ is the solution to the energy equation of Eq.~(\ref{BIC}), {that is, a discrete energy level.} In Fig.~\ref{cc}, we denote the corresponding energy level by the blue line. This state is actually the BIC, which can be verified both numerically and analytically as follows.

With the assistance of the inverse Fourier transformation
{\begin{eqnarray}
{\beta_j}=\frac{1}{\sqrt{N_c}}\sum_k\beta_k e^{-ikj},
\end{eqnarray}}
we can obtain the photonic distribution for the BIC ($E=\Omega$) as (see Appendix~\ref{appB})
\begin{equation}
\label{BICp}
{\frac{\beta_{j}}{\alpha}}= \begin{array}{l}
  \left\{\begin{matrix}
   0 &,&  j< 0 \& j>  N \\
  \frac{2g\sin Kj}{\sqrt{4\xi^2-\Omega^2}}&,&  0\le j\le N
\end{matrix}\right.
\end{array}
\end{equation}
It means that, the photon in the waveguide is trapped within the regime which is covered by the giant atom, and therefore we name it as BIC. The BIC is different from other states in the continuum, in which the photon distributes along the whole waveguide.

In Figs.~\ref{bshape} (c) and (d), we demonstrate the standing wave type photonic distribution for the BIC.  Here, the solid bar represents the numerical results which come from the direct diagonalization of the Hamiltonian in the single excitation subspace, while the empty circles are the analytical results based on Eq.~(\ref{BICp}).  For the parameter $\Omega/\xi=-1$, we will have $K=\pi/3$. The distribution in Fig.~\ref{bshape} (c) shows that the photon occupies at the $j=1,2,4$ and $5$th sites, but not at $j=3$.  On the contrary, for the parameter of $\Omega/\xi=-\sqrt{2},\phi=0,N=12$, there is no photon distribution between the two legs of the giant atom at the site with $j=4m\,(m=0,1,2,3)$. Therefore, {the frequency} and the photonic distribution of the BIC can be modulated by the size and frequency of the giant atom as well as its coupling phase to the waveguide.

\section{BIC-BOC transition dynamics}

In this section, we will investigate the dynamics of the system for both of the atomic and photonic counterpart under the parameter of $\Omega=-\xi, N=6, \phi=\pi$, in which case the energy spectrum is given in Fig.~\ref{cc}(a).

\begin{figure}
  \centering
  \includegraphics[width=4cm]{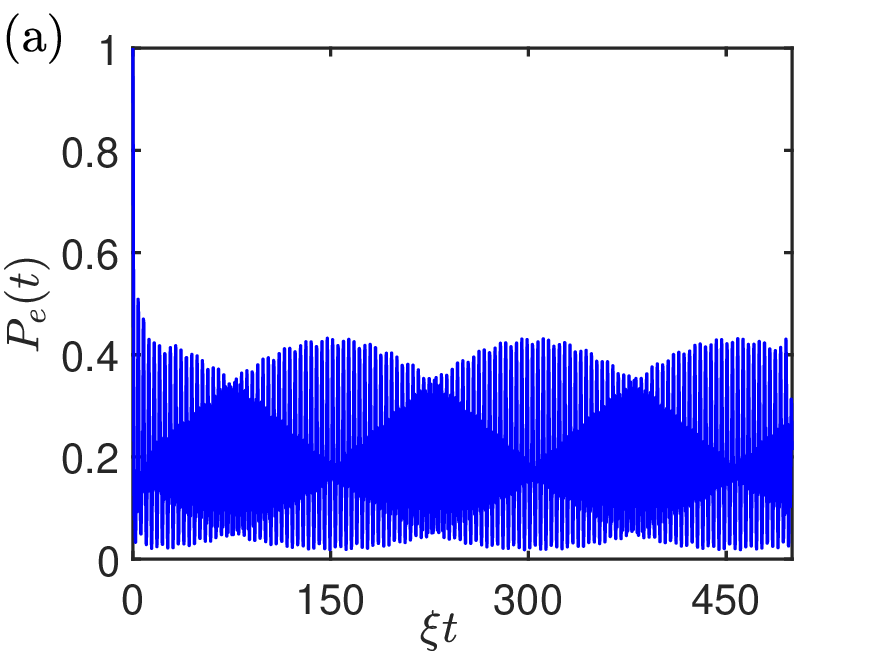}\includegraphics[width=4cm]{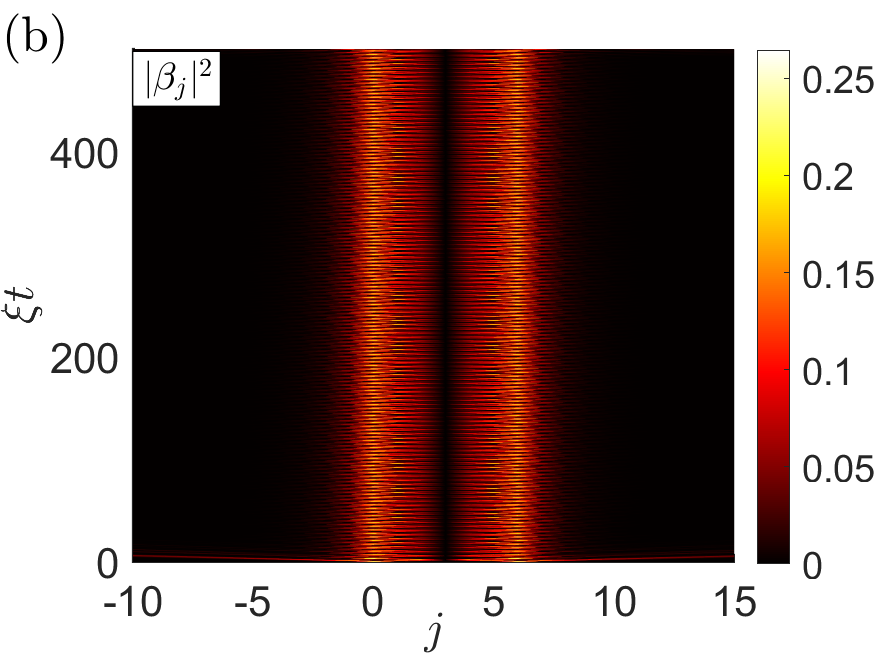}
  \includegraphics[width=4cm]{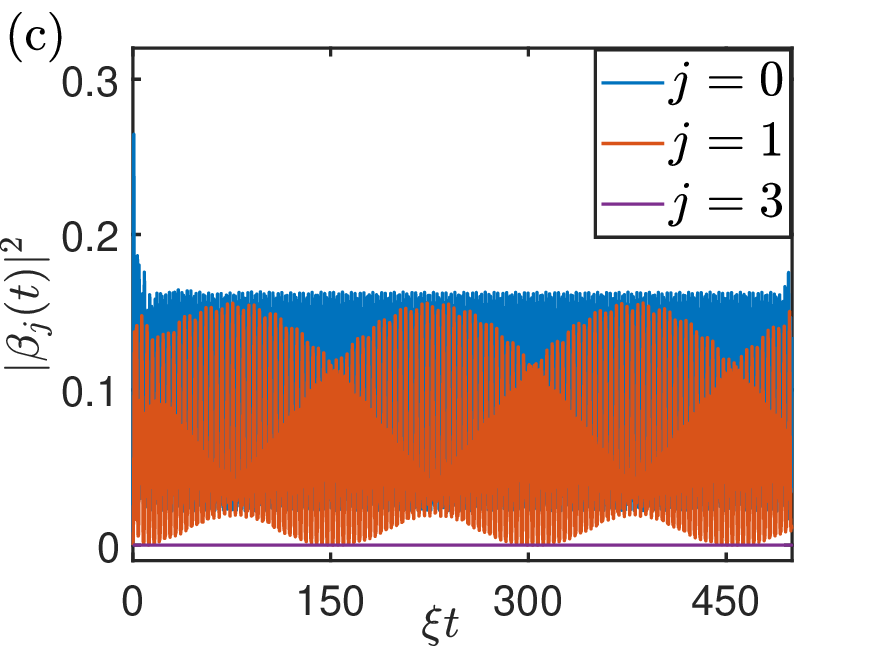}\includegraphics[width=4cm]{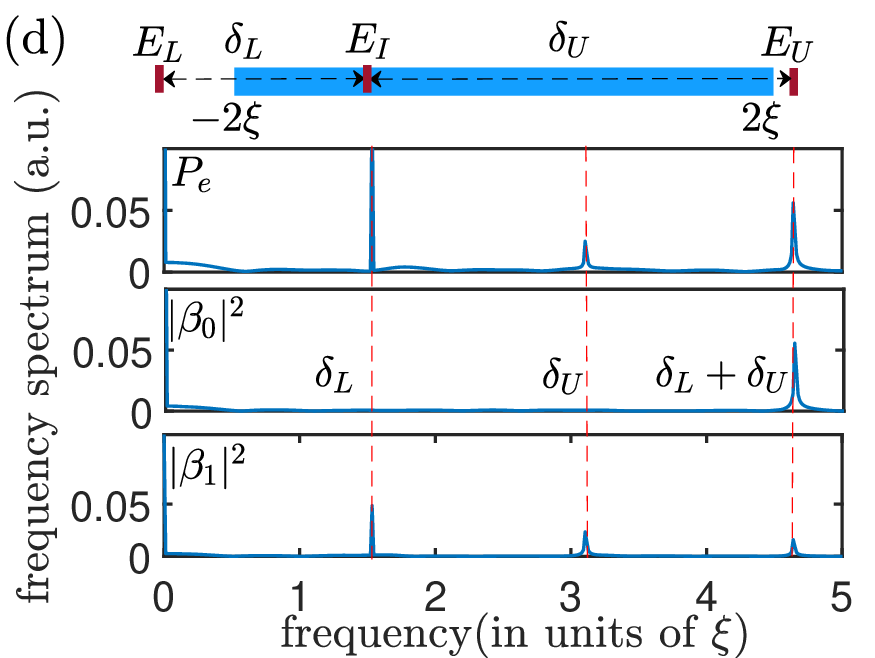}
  \caption{The dynamical evolution of the system. (a) The dynamics of the atomic population in its excited states. (b,c) The dynamics of the photonic distribution. (d) The frequency spectrum of the atomic and photonic dynamical evolution which us obtained by the numerical Fourier transformation from the time domain to the frequency domain. The parameters are set as $\omega_c=0,\Omega=-\xi, g=1.1\xi, N=6,\phi=\pi$.}\label{pai}
\end{figure}

We set the initial state of the system as $|\psi(0)\rangle=\sigma_+|G\rangle$, in which the giant atom is in its excited state while all of the resonators are in their ground states. In Fig.~\ref{pai}, we numerically illustrate the dynamical evolution of the system, which is governed by $|\psi(t)\rangle=\exp(-iHt)|\psi(0)\rangle$. In Fig.~\ref{pai}(a), we can observe that $P_e=|\langle \psi(t)|e\rangle|^2$ shows a slow variable envelope and the fast oscillation in each of the envelope. The dynamics of the photonic counterpart $\beta_j=\langle \psi(t)|a_j^\dagger |G\rangle$ is shown in Fig.~\ref{pai}(b).  The photon emitted by the giant atom is nearly completely trapped inside the covered regime. We further detailed plot the evolution of the photonic distribution in the resonator $j=0,1,3$ in Fig.~\ref{pai}(c). The photon can never be excited in the $3$th resonator, that is $|\beta_3|=0$ during the time evolution. For $j=1$, we find the similar quantum beats phenomenon with the atomic dynamics and the same result for $j=2$ is not shown in the figure.  For $j=0$, a single high oscillation frequency is observed.

To understand the above dynamical process, we write the initial state as
\begin{eqnarray}
|\psi(0)\rangle= c_U |\phi_U\rangle+c_I |\phi_I\rangle+c_L |\phi_L\rangle+\sum_k c_k
|\phi_k\rangle,
\end{eqnarray}
where $|\phi_U\rangle$ and $|\phi_L\rangle$ are the upper and lower BOCs with eigenenergy $E_U$ and $E_L$  respectively, and $|\phi_I\rangle$ is the BIC with eigenenergy $E_I$. $|\phi_k\rangle$ is the $k$th propagating photonic mode in the waveguide, and the $C$-numbers $c_m\,(m=U,I,L,k)$ denote the corresponding amplitudes. Extracting from the exact numerical results in Fig.~\ref{cc}(a), the eigenenergy of the system is illustrated in the cartoon figure in the upper panel
of Fig.~\ref{pai}(d). Here, the continual band, which locates in the regime of $[-2\xi,2\xi]$ is demonstrated by the horizontal blue bars, while the BIC and the BOCs are represented by the vertical dark red bar. It is well known that the propagating modes will not contribute to the atomic dynamics when the evolution time is long enough due to the interference effect~\cite{ZZ,bicAN}. Therefore, in the long time limit, we will have
\begin{eqnarray}
P_e(t)&=& |e^{-iE_Lt}c_L^2+e^{-iE_It}c_I^2+e^{-iE_Ut}c_U^2|^2\nonumber\\
&=& |e^{i\delta_Lt}c_L^2+c_I^2+e^{-i\delta_Ut}c_U^2|^2,\label{paie}\\
|\beta_j(t)|^2&=&|e^{-i\delta_Ut}c_U d_{U,j}+e^{i\delta_Lt}c_L d_{L,j}+c_I d_{I,j}|^2.\label{paiew}
\end{eqnarray}
where $\delta_{L}=E_I-E_L$ and $\delta_{U}=E_U-E_I$ are the detunings between of the BOCs and the BIC, and $d_{\alpha,j}=\langle \phi_\alpha|a_j^\dagger|G\rangle\,(\alpha=U,I,L)$ denotes the photonic excitation amplitude on the $j$th resonator of the waveguide when the system is in the bound state $|\phi_\alpha\rangle$.

To further investigate the dynamical property of the system, we apply the numerical fast Fourier transformation technique to obtain the frequency spectrum for $P_e$ and $|\beta_j|^2\,(j=0,1)$ in the three lower panels of Fig.~\ref{pai}(d). We can observe  three significant peaks for $P_e$ and $|\beta_1|^2$, which locate at the frequency $\delta_L$, $\delta_U$ and $\delta_U+\delta_L$, agreeing with the analytical expression in Eqs.~(\ref{paie}) and (\ref{paiew}). It implies that the dynamics of the giant atom is induced by the oscillations between the single BIC and the two BOCs (which contributes the peaks located at $\delta_U$ and $\delta_L$) and between the two BOCs (which contributes the peaks located at $\delta_U+\delta_L$).
On the contrary, we only find a single peak which locates at the frequency of $\delta_U+\delta_L$ in the spectrum of $|\beta_0|^2$. This can also be explained by the analytical expression in Eq.~(\ref{paiew}), in which there only exists a single non-zero frequency component at $\delta_U+\delta_L$ since $d_{I,0}=0$ as shown in Fig.~\ref{bshape}(c). It further indicates that the oscillation between the two BOCs contributes to the photonic dynamics in the zeroth cavity.

\section{Conclusions}

In this paper, we have studied the engineering of the BOCs and BIC in a one-dimensional waveguide QED setup with the giant atom. We find that the number of the BOCs can be controlled on demand by the size of the giant atom and its coupling phase to the waveguide. We also prove that the BIC is exactly resonant with the frequency of the giant atom.  Induced by the BIC-BOC oscillation, both of the atomic and photonic dynamics are characterized by a quantum beat phenomenon.

The giant atom can be realized by the superconducting transmon qubit, and the coupled resonator waveguide up to tens of sites~\cite{22prx,CRW2} has also been fabricated in superconducting circuits. In these experimental realizations, the parameters can be achieved in the regime of $g\ll\xi\approx 100-200$ MHz, and therefore our predicted dynamical behavior can be observed even when the giant atom suffers spontaneous emission with the life time of $T_1=10 {\rm \mu s}$~\cite{matter2020}.

Beyond the specific setting considered in this work, our investigation can be developed to the system consisting of more giant atoms or one giant atom which couples to the waveguide via more than two sites with the on-demand coupling strengths and phases. In these more complicated setups, the BIC and BOC is deserved to be deeply investigated  and is hopefully to design the needed dynamical process for quantum information processing.

\begin{acknowledgments}
This work is supported by the funding from Jilin Province (Grant Nos. 20230101357JC and 20220502002GH) and the National Science Foundation of China (Grant No. 12375010).
\end{acknowledgments}

\appendix
\addcontentsline{toc}{section}{Appendices}\markboth{APPENDICES}{}
\section{Bound state out of the continuum}
\label{ABOC}
{In Eqs.~(\ref{BIC},\ref{EUm},\ref{ELm}) of the main text, we have given the transcendental equations for the energies of the BOCs. In this appendix, we will give a detailed derivation.}

We begin with the Hamiltonian in Eq.~(\ref{Hk}) and the wave function in Eq.~(\ref{wavek}).  The Sch\"{o}dinger equation $H|\phi\rangle=E|\phi\rangle$ yields
\begin{eqnarray}
&&\alpha (E-\Omega)=\frac{g}{\sqrt{N_c}}\sum_k \beta_k[1+e^{-i(kN+\phi)}],\nonumber\\
&&(E+2\xi \cos k)\beta_k=\frac{g\alpha}{\sqrt{N_c}}[1+e^{i(kN+\phi)}].\label{eigf}
\end{eqnarray}

By eliminating $\alpha$ and $\beta_k$, the equation for the eigen energy $E$ can be obtained as
\begin{eqnarray}
E-\Omega&=&Y(E)\equiv\frac{g^2}{\pi}\int_{-\pi}^{\pi}\frac{1+\cos(kN+\phi)}{E+2\xi\cos k}dk,
\end{eqnarray}
which is Eq.~(\ref{BIC}) in the main text. Now, we perform the integral to calculate $Y(E)$ by introducing a new variable $z=e^{ik}$. We can achieve

\begin{eqnarray}
Y(E)&=&\frac{g^2}{\pi}\int_{-\pi}^{\pi}\frac{1+\cos(kN+\phi)}{E+2\xi\cos k}dk \nonumber\\
&=&\frac{g^2}{2\pi}\int_{-\pi}^{\pi}\frac{2+e^{i(kN+\phi)}+e^{i(kN-\phi)}}{E+\xi(e^{ik}+e^{-ik})}dk\nonumber\\
&=&\frac{g^2}{2i\pi\xi}\oint_{|z|=1}\frac{2+z^Ne^{i\phi}+z^Ne^{-i\phi}}{z^2+E z/\xi+1}dz\nonumber\\
&=&\frac{g^2}{2i\pi\xi}\oint_{|z|=1}\frac{2+z^Ne^{i\phi}+z^Ne^{-i\phi}}{(z-z_1)(z-z_2)}dz
\end{eqnarray}
where $z_1=-(E/2\xi)+\sqrt{(E/2\xi)^2-1}$ and $z_2=-(E/2\xi)-\sqrt{(E/2\xi)^2-1}$. In the regime of $|E|>2\xi$, we will have
\begin{eqnarray}
Y(E_U)&=&\frac{g^2}{\xi}\frac{2+2 z_1^N\cos \phi}{z_1-z_2}\nonumber\\
&=&\frac{g^2}{\xi}\frac{1+[-\frac{E_U}{2\xi}+
\sqrt{(\frac{E_U}{2\xi})^2-1}]^N\cos\phi}{\sqrt{(\frac{E_U}{2\xi})^2-1}},\label{EU}\\
Y(E_L)&=&\frac{g^2}{\xi}\frac{2+2 z_2^N\cos \phi}{z_2-z_1}\nonumber\\
&=&-\frac{g^2}{\xi}\frac{1+[-\frac{E_L}{2\xi}
-\sqrt{(\frac{E_L}{2\xi})^2-1}]^N\cos\phi}{\sqrt{(\frac{E_L}{2\xi})^2-1}}.\label{EL}
\end{eqnarray}
Here, we have assumed $E_U>2\xi$ and $E_L<-2\xi$, which represent the upper and lower BOCs, respectively. Therefore, we obtain Eqs.~(\ref{EUm}) and (\ref{ELm}) in the main text.

\section{Bound state in the continuum}
\label{appB}

In this appendix, we will give the detailed derivation of the energy and the wave function of the BIC.

Firstly, we  will prove that $E=\Omega$ is a solution to the transcendental equation (\ref{BIC}) in the main text. That is to prove
\begin{eqnarray}
Y(\Omega)=\frac{g^2}{\pi}\int_{-\pi}^{\pi}\frac{1+\cos(kN+\phi)}{\Omega+2\xi\cos k}dk \label{1}
\end{eqnarray}
is always zero. To this end, we rewrite $Y(\Omega)$ as
\begin{eqnarray}
Y(\Omega)&=&\frac{g^2}{\pi}\int_{-\pi}^{\pi}\frac{1+\cos(kN+\phi)}{\Omega+2\xi\cos k}dk\nonumber\\
&=&\frac{g^2}{2\pi}\int_{-\pi}^{\pi}\frac{2+e^{i(kN+\phi)}+e^{-i(kN+\phi)}}{\Omega+\xi(e^{ik}+e^{-ik})}dk\nonumber\\
&=&\frac{g^2}{2i\xi \pi}\oint_{|z|=1}\frac{2z^N+z^{2N}e^{i\phi}+e^{-i\phi}}{(z-z_1)(z-z_2)z^N}dz,
\end{eqnarray}
where $z_{1,2}=-(\Omega/2\xi)\pm i\sqrt{1-(\Omega/2\xi)^2}=e^{\pm iK}$ which satisfies $|z_{1,2}|=1$.
Applying the residue theorem to solve the above integral, only the $N$-order singular point~$(z=0)$ is needed to be considered.
Therefore,
\begin{eqnarray}
Y(\Omega)&=&\frac{g^2}{\xi}\frac{1}{(N-1)!}\frac{d^{N-1}}{dz^{N-1}}[\frac{2z^N+z^{2N}e^{i\phi}+e^{-i\phi}}{(z-z_1)(z-z_2)}]|_{z=0}\nonumber\\
&=&\frac{g^2 e^{-i\phi}}{2i\xi\sqrt{1-(\Omega/2\xi)^2}}(z_2^{-N}-z_1^{-N})\nonumber\\
&=&\frac{g^2 }{2i\xi\sqrt{1-(\Omega/2\xi)^2}}(e^{-i(KN+\phi)}-e^{i(KN-\phi)})\nonumber\\
&=&0,
\end{eqnarray}
where we have used the condition $G(K)=G(-K)=0$ for the BIC.

Next, we will study the wave function of the BIC.  From Eq.~(\ref{eigf}), we will have
{\begin{eqnarray}
\frac{\beta_k}{\alpha}&=&\frac{g}{\sqrt{N_c}}\frac{1+e^{i(kN+\phi)}}{E+2\xi \cos k}.
\end{eqnarray}}
Going back into the real space, we will further obtain
{\begin{eqnarray}
\frac{\beta_j}{\alpha}&=&\frac{g}{N_c}\sum_k \frac{[1+e^{i(kN+\phi)}]e^{-ikj}}{\Omega+2\xi\cos k}\nonumber\\
&=&\frac{g}{2\pi}\int_{-\pi}^{\pi}\frac{[1+e^{i(kN+\phi)}]e^{-ikj}}{\Omega+2\xi\cos k}dk\nonumber\\
&=&\frac{g}{2i\pi\xi}\oint_{|z|=1}\frac{1+z^Ne^{i\phi}}{(z-z_1)(z-z_2)z^j}dz.
\end{eqnarray}}
When $j<0$, there is no singular point in the integral domain, so
\begin{eqnarray}
\frac{\beta_j}{\alpha}=\frac{g}{2i\pi\xi}\oint_{|z|=1}\frac{1+z^Ne^{i\phi}}{(z-z_1)(z-z_2)z^j}dz=0.
\end{eqnarray}
When $0\leq j\leq N $,
\begin{equation}
\oint_{|z|=1}\frac{z^Ne^{i\phi}}{(z-z_1)(z-z_2)z^j}dz=0.
\end{equation}
Therefore, we will achieve
\begin{eqnarray}
\frac{\beta_j}{\alpha}&=&\frac{g}{2i\pi\xi}\oint_{|z|=1}\frac{1}{(z-z_1)(z-z_2)z^j}dz\nonumber\\
&=&\frac{g}{\xi(j-1)!}\frac{d^{j-1}}{dz^{j-1}}[\frac{1}{(z-z_1)(z-z_2)}]|_{z=0}\nonumber\\
&=&\frac{g }{2i\xi\sqrt{1-(\Omega/2\xi)^2}}(z_1^j-z_2^j)\nonumber\\
&=&\frac{g }{2i\xi\sqrt{1-(\Omega/2\xi)^2}}(e^{iKj}-e^{-iKj})\nonumber\\
&=&\frac{g\sin(Kj) }{\xi\sqrt{1-(\Omega/2\xi)^2}}.
\end{eqnarray}

When $j> N$
\begin{eqnarray}
\frac{\beta_j}{\alpha}&=&\frac{g}{\xi(j-1)!}\frac{d^{j-1}}{dz^{j-1}}[\frac{1}{(z-z_1)(z-z_2)}]|_{z=0}\nonumber\\
&&+\frac{g}{\xi(j-N-1)!}\frac{d^{j-N-1}}{dz^{j-N-1}}[\frac{e^{i\phi}}{(z-z_1)(z-z_2)}]|_{z=0}\nonumber\\
&=&\frac{g }{2i\xi\sqrt{1-(\Omega/2\xi)^2}}(z_2^{-j}-z_1^{-j}+z_2^{-(j-N)}-z_1^{-(j-N)})\nonumber\\
&=&\frac{ge^{iKj} }{2i\xi\sqrt{1-(\Omega/2\xi)^2}}(1+e^{i(KN+\phi)})\nonumber\\
&&-\frac{ge^{-iKj} }{2i\xi\sqrt{1-(\Omega/2\xi)^2}}(1+e^{i(-KN+\phi)})\nonumber\\
&=&0.
\end{eqnarray}
Up to now, we have obtained the photonic distribution in BIC which is given in Eq.~(\ref{BICp}) in the main text.

\end{document}